**Ibraheem Kasim Ibraheem[1,*]**

**Regular paper**

# Adaptive System Identification Using LMS Algorithm Integrated with Evolutionary Computation

System identification is an exceptionally expansive topic and of remarkable significance in the discipline of signal processing and communication. Our goal in this paper is to show how simple adaptive FIR and IIR filters can be used in system modeling and demonstrating the application of adaptive system identification. The main objective of our research is to study the LMS algorithm and its improvement by the genetic search approach, namely, LMS-GA, to search the multi-modal error surface of the adaptive IIR filter to avoid local minima and finding the optimal weight vector when only measured or estimated data are available. Convergence analysis of the LMS algorithm in the case of the colored input signal, i.e., the correlated input signal is demonstrated on adaptive FIR filter via the input's power spectral density and the Fourier transform of the autocorrelation matrix of the input signal. Simulations have been carried out on adaptive filtering of FIR and IIR filters and tested on white and colored input signals to validate the powerfulness of the genetic-based LMS algorithm.

Keywords: System identification, LMS algorithm, adaptive filtering, genetic algorithm, colored signals, power spectral density, multi-modal error surface.

## 1. Introduction

Adaptive filters are systems whose structure is alterable or adjustable in such a way that its behavior or performance improves through contact with its environment. Such systems usually can automatically adapt in the face of changing environments, they can be trained to achieve particular filtering, and they do not require elaborate synthesis procedures usually needed for non-adaptive systems, other characteristics can be found in [1].

Traditional non-adaptive filters which are utilized for extraction of data from a certain input sequence have typically the linearity and time-invariance properties. While for the case of the adaptive filters, the limitation of invariance is eliminated. This is achieved by enabling the filter to update its own weights as per certain foreordained optimization process. Adaptive digital filters can be classified into adaptive Finite Impulse Response (FIR) filter, or commonly known as an *Adaptive Linear Combiner* which is unconditionally stable, and Infinite Impulse Response (IIR) presents a prospective enhancement in the performance and less computation power than corresponding adaptive FIR filter.

Common applications of adaptive filters are noise cancelation, inverse modeling, prediction, jammer suppression [2]–[4], and system identification, which is the main topic of this paper.

Adaptive system identification had a long history of many types of research ranged from the implementation of neural networks [5]–[9] to swarm optimization algorithms [10]–[14], reaching to the application of LMS adaptation algorithm on IIR and FIR adaptive filters on different applications [3], [15]–[18]. Application of genetic algorithm and its variant in system identification are studied in [4], [19] respectively.

The major drawback with the standard LMS algorithm in system identification is that, the adaptive IIR digital filter suffers from the multimodality of the error surface versus the filter

[1*] Corresponding author: Ibraheem Kasim Ibraheem, Electrical Engineering Department, College of Engineering, Baghdad University, E-mail: ibraheemki@coeng.uobaghdad.edu.iq

coefficients, and it is easy that the adaptation techniques (e.g., standard LMS algorithm) get stuck at one of the local minima and diverge away from the global optimum solution. The global minimum of the error surface is found in the LMS algorithm by traveling toward the negative direction of the error gradient. In the case of a multi-modal error surface, the LMS algorithm like the vast majority of the learning techniques may drive the filter into a local minimum. Moreover, the initial choice of the filter coefficients and the proper selection of the step size particularly determine the convergence behavior of the LMS algorithm [4].

An evolutionary algorithm named Genetic Algorithm (GA) is presented for multi-modal error surface searching in IIR adaptive filtering [4]. Nevertheless, the high computational complexity and slow convergence are the main drawbacks of utilizing such an algorithm. Started by the benefits and deficiencies of the evolutionary algorithm and gradient descent algorithm, we build up a novel integrated searching algorithm, namely, LMS-GA, where the GA searching algorithm is integrated with the standard LMS algorithm. The proposed LMS-GA algorithm has the attributes of simple implementation, global searching ability, rapid convergence, and less sensitivity to the parameters selection.

*Paper Findings*. This paper reviews the implementation of LMS algorithm in the adaptation of FIR digital filters with an application on system identification discusses the effect of colored input signals on the convergence rate of the adaptation process. Furthermore, developing a new search algorithm, namely, LMS-GA for learning adaptive IIR digital filters coefficients using the gradient descent algorithm integrated with the evolutionary computations. The algorithm is designed in such a way that as soon as the adaptive IIR filter is found to have a sluggish convergence or to be trapped at a local minimum, the adaptive IIR digital filter parameters are updated in a random behavior to move away from the local minimum and possess a higher chance of traveling toward the global optimum solution.

The current paper is structured as follows: Section 2 presents the motivation to adopt the new LMS-GA learning algorithm for adaptation of IIR and FIR digital filters. The basic structures of FIR and IIR adaptive digital filters are given in Section 3. Application of LMS algorithm on both adaptive FIR and IIR digital filters is demonstrated in Section 4. A concise overview of GA is introduced in Section 5. The main results are presented in Section 6, it includes the discussion of the effect of the colored input signal on the adaptation process and investigating the new LMS-GA learning technique with its application as a learning tool. The numerical results are contained in Section 7. Finally, the paper is concluded in Section 8.

## 2. Motivation

Usually, gradient descent algorithms can only do well locally; whereas GA is reasonably sluggish to "calibrate" the optimal solution once a fitting region in the searching space is found. Furthermore, The GA likewise necessitates calculating the values of the fitness function for all the chromosomes in the population which render it an algorithm with a high computational complexity. We offer a novel learning technique for the adaptation of adaptive FIR and IIR filters to cope with the difficulties of GA and gradient descent techniques, namely, LMS-GA. This new learning tool incorporates the quintessence and features of both algorithms.

## 3. Preliminaries on Adaptive FIR and IIR Filtering

The FIR filter is shown in Fig. 1 in the form of a single-input transversal filter. The "adaptation" is that by which the weights are adjusted or adapted in response to a function of the error signal. When the weights are in the process of being adjusted, they, too, are a function of the input components and not just the output so that the latter is no longer a linear function of the input.

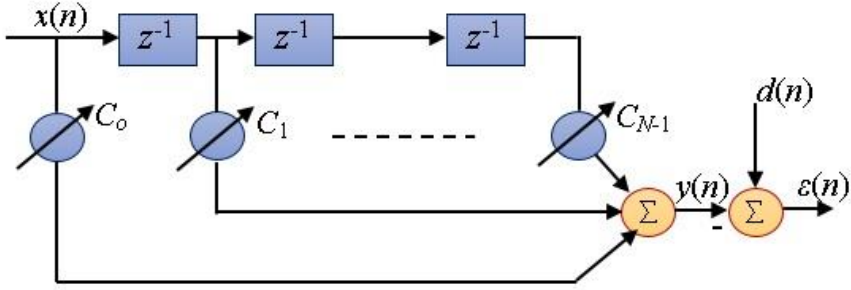

Fig. 1. Adaptive FIR filter Implemented as Transversal Filter [1].

$$y(n) = \sum_{k=0}^{N-1} w_k(n)\, x(n-k) \tag{1}$$

where $N$ is the order of the filter. In vector notation:

$$y(n) = X_N^T(n)\, C_N(n) \tag{2}$$

and the error signal is given as

$$\varepsilon(n) = d(n) - y(n) = d(n) - X_N^T(n)\, C_N(n) \tag{3}$$

where $d(n)$ is the desired signal, $y(n)$ is the filter output, $X_N^T(n)$ is the input signal vector, $X_N^T(n) = [x(n)\ x(n-1) \cdots x(n-N+1)]$, $C_N(n) = [c_o(n)\ c_1(n) \cdots c_{N-1}(n)]$ is the coefficient vector, $n$ representing the time index, the superscript $T$ denoting transpose operator and the subscript $N$ representing the dimension of a vector.

**Remark**: To a set of points of the inputs sequence $x(n)$ and the reference waveform $d(n)$, there corresponds an optimum coefficient vector or impulse response $C_N(n)$. Given another set of points, there is no guarantee that the resulting optimum vector is related to the first unless the properties of the waveform do not change over different sections. Based on this, we can formulate the following assumption.

**Assumption (H1)**: the input sequence $x(n)$ and the reference waveform $d(n)$ are stochastic processes. Then the error $\varepsilon(n)$ defined by (3) is also stochastic.

The performance function or the mean-square error $E_{ms}$ is defined as,

$$E_{ms} = E\{|\varepsilon(n)|^2\} \tag{4}$$

Then,

$$E_{ms} = E\{|d(n) - y(n)|^2\}$$

Expanding and using (2) we obtain [20],

$$E_{ms} = E\{|d(n)|^2\} - C_N^T Re[E\{d^*(n)X_N(n)\}] + C_N^T E\{X_N(n)X_N'(n)\}C_N^*$$

where $*$ and $'$ denote conjugate and conjugate transpose respectively and the time index n is omitted from $C_N$ for simplicity. If we define the expected value of $d(n)$ as

$$D_{ms} \triangleq E\{|d(n)|^2\} \tag{5}$$

and the ensemble or statistical autocorrelation matrix $R_{ms}$ of $x(n)$ as,

$$R_{ms} \triangleq E\{X_N(n)X_N'(n)\} \tag{6}$$

which is a Toeplitz matrix and the ensemble average cross-correlation vector as,

$$P_{ms} \triangleq E\{d^*(n)X_N(n)\} \tag{7}$$

Then, $E_{ms}$ can be written as [20],

$$E_{ms} = D_{ms} - 2C_N^T Re\{P_{ms}\} + C_N^T R_{ms} C_N^* \tag{8}$$

which shows that $E_{ms}$ has a quadratic form. To find the choice for $C_N$ that minimizes $E_{ms}$, we find the gradient of $E_{ms}$ w.r.t. $C_N$ and find the optimum value of the weights $C_{ms}^o$ which sets it to zero. This leads to [20],

$$R_{ms} C_{ms}^o = Re\{P_{ms}\}$$

The solution is unique if $R_{ms}$ is invertible, and then

$$C_{ms}^o = R_{ms}^{-1} Re\{P_{ms}\} \tag{9}$$

which is the Wiener-Hopf equation in matrix form. Recursive filters like IIR with poles as well as zeros would offer the same advantages (resonance, sharper cut off, ...etc.) that non-recursive filter offers in time-invariant applications. The recursive filters have two main weakness points, they become unstable if the poles move outside the unit circle and their performance indices are generally non-quadratic and may even have a local minimum. The adaptive IIR filter may be represented in the standard adaptive model as illustrated in Fig. 2.

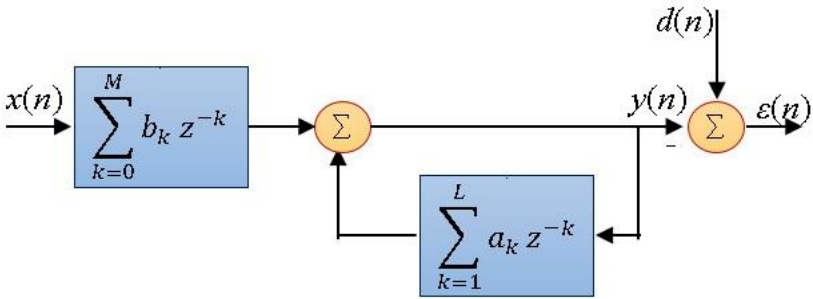

Fig. 2. Recursive adaptive IIR.

The input-output relationship is expressed as,

$$y(n) = \sum_{k=0}^{M} b_k x(n-k) + \sum_{k=1}^{L} a_k y(n-k) \tag{10}$$

where $b_k$'s and $a_k$'s are the coefficients of the IIR filter, and $x(n)$ and $y(n)$ are the input and output of the IIR filter respectively. The transfer function for the IIR filter is given by [1],

$$H(z) = \frac{Y(z)}{X(z)} = \frac{b_0 + b_1 z^{-1} + \cdots + b_M z^{-1}}{1 - a_1 z^{-1} + \cdots + a_L z^{-1}} = \frac{\sum_{k=0}^{M} b_k z^{-k}}{1 - \sum_{k=1}^{L} b_k z^{-k}}$$

Note that in (10) the current output sample is a function of the past output $y(n-k)$, as well as the present and past input sample $x(n)$, and $x(n-k)$, respectively. The strength of the IIR filter comes from the flexibility the feedback arrangement provides. For example, an IIR filter normally requires fewer coefficients than FIR filter for the same set of specifications.

## 5. System Identification Using Adaptive Fir and IIR Filters With LMS Algorithm

Newton's and steepest descent methods are used for descending toward the minimum on the performance surface. Both require an estimation of the gradient in each iteration. The gradient estimation method is general because they are based on taking differences between estimated points on the performance surface, that is, the difference between estimates of the error $\varepsilon(n)$. In this section, we will use another algorithm for descending on the performance surface, known as Least Mean Square (LMS) algorithm and will be investigated on both FIR and IIR digital filters.

5.1. The LMS algorithm and Adaptive FIR Filtering

Given a record of data $(x(n), d(n))$, one can compute $R_{ms}$ and $P_{ms}$, $R_{ms}$ might not be invertible and if it is, its inversion requires high numerical precision. A method depending on search techniques has the advantage of being simple to implement but at the expense of some inaccuracy in the final estimate. We use the gradient or steepest descent searching technique to find $C_{ms}^o$ iteratively. This technique is applicable to the minimization of the quadratic performance function $E_{ms}$, since it is a convex function of the coefficients $C_N$, i.e., it possesses global minimum. A gradient vector is computed as $\partial E_{ms}/\partial C_N$, $i = 1, \cdots, N-1$, at this point and each tap weight is changed in the direction opposite to its corresponding gradient component and by an amount proportional to the size of that component. Therefore,

$$C_N(l+1) = C_N(l) - \mu \nabla_{C_N} E_{ms} \tag{11}$$

where $\nabla_C E_{ms}$ is the gradient vector, the subscript indicating that the gradient is taken w.r.t. to the components of the coefficient vector $C_N$, $l$ is the iteration number, and $\mu$ is the convergence factor that regulates the speed and the stability of the adaptation. It is clear from Fig. 3 which represents the one-dimensional case how repeating this procedure leads to the minimum of $E_{ms}$ and hence the optimal value $C_{ms}^o$. Approximating the gradient of $E_{ms}$ by the gradient of the instantaneous squared error, i.e.,

$$\nabla_{C_N} E_{ms} = \nabla_{C_N} E\{|\varepsilon(n)|^2\} \approx \nabla_{C_N} |\varepsilon(n)|^2 = 2\varepsilon(n) \cdot \frac{d\varepsilon(n)}{d\,C_N} = 2\varepsilon(n) X_N(n)$$

we may write

$$C_N(l+1) = C_N(l) - \mu \nabla_{C_N} |\varepsilon(n)|^2 = C_N(l) + 2\mu\,\varepsilon(n)\,X_N(n) \tag{12}$$

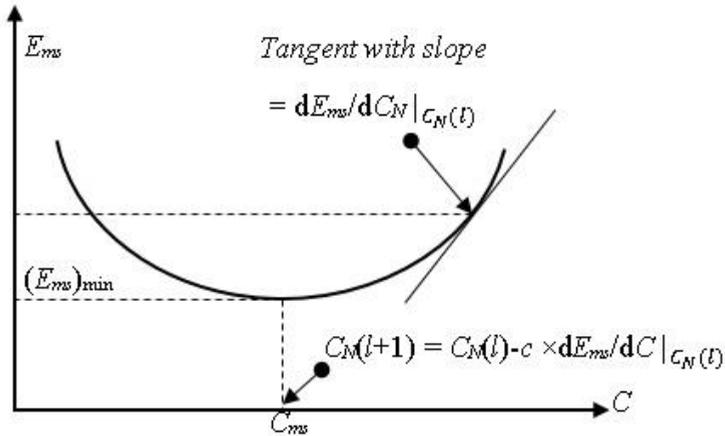

Fig. 3. Gradient descent on a one-dimensional projection of a quadratic performance function.

Usually, $C_N(l)$ is updated for every sample as is the case when variations are to be tracked in an estimation process. When this is true, $l = n$. The LMS algorithm is summarized as follows:

$$\begin{cases} y(n) = C^T(n-1)X_N(n) \\ \varepsilon(n) = d(n) - y(n) \\ C_N(n) = C_N(n-1) + 2\mu\,\varepsilon(n) X_N(n) \end{cases} \tag{13}$$

A flowchart for the LMS algorithm is given in Fig.4. A convergence analysis for the LMS algorithm has been done in [1], [20] and concluded that to achieve convergence the value of $\mu$ is found as,

$$0 < \mu < \frac{1}{\lambda_{max}} \tag{14}$$

where $\lambda_{max}$ is the maximum eigenvalue of $R_{ms}$.

5.2. Adaptation of IIR digital filter Based on LMS algorithm

To develop an algorithm for the recursive IIR filter, let us define the time-varying vector $C_N(n)$ and the signal $U(n)$ as follows,

$$C_N(n) = [b_o(n)\ b_1(n) \cdots \cdots b_M(n)\ a_1(n) \cdots \cdots a_L(n)] \tag{15}$$
$$U(n) = [x(n)\ x(n-1) \cdots x(n-M)\ y(n-1) \cdots \cdots y(n-L)] \tag{16}$$

From Fig. 2 and equation (10), we can write
$$\varepsilon(n) = d(n) - y(n) = d(n) - C_N(n)^T U(n) \tag{17}$$

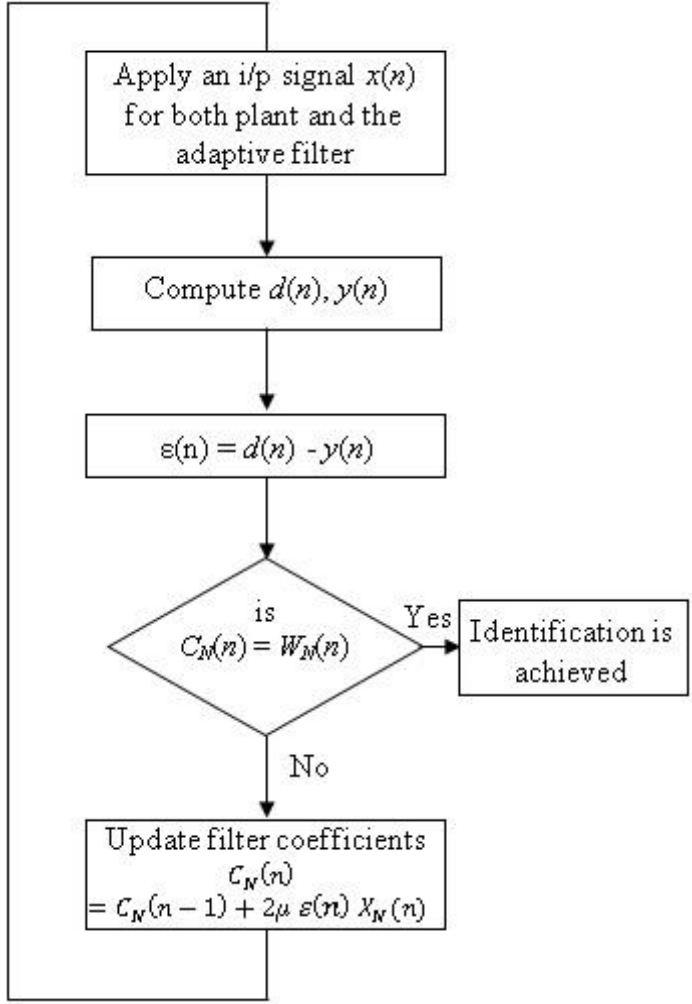

Fig. 4. Flowchart of LMS algorithm.

This is quite similar to the non-recursive case in (3). The main difference being that $U(n)$ contains values of $y(n)$ as well as $x(n)$. We again use the gradient approximation,

$$\nabla_{C_N} E_{ms} = \nabla_{C_N} E\{|\varepsilon(n)|^2\} \approx \frac{\partial \varepsilon^2}{\partial C_N(n)} = 2\varepsilon \frac{\partial \varepsilon}{\partial C_N(n)}$$
$$= 2\varepsilon \left[\frac{\partial \varepsilon(n)}{\partial b_o(n)} \cdots \frac{\partial \varepsilon(n)}{\partial b_M(n)} \frac{\partial \varepsilon(n)}{\partial a_1(n)} \cdots \frac{\partial \varepsilon(n)}{\partial a_L(n)}\right]^T$$
$$= -2\varepsilon(n) \left[\frac{\partial y(n)}{\partial b_o(n)} \cdots \frac{\partial y(n)}{\partial b_M(n)} \frac{\partial y(n)}{\partial a_1(n)} \cdots \frac{\partial y(n)}{\partial a_L(n)}\right]^T \quad (18)$$

The derivatives in (18) present a special problem because $y(n)$ is now a recursive function. Using (10) we define

$$\alpha_k(n) \triangleq \frac{\partial y(n)}{\partial b_k(n)} = x(n-k) + \sum_{l=1}^{L} a_l \frac{\partial y(n-l)}{\partial b_k(n)}$$
$$= x(n-k) + \sum_{l=1}^{L} a_l \alpha_k(n-l), \quad k = 0,1,\cdots,M \quad (19)$$

$$\beta_k(n) \triangleq \frac{\partial y(n)}{\partial a(n)} = y(n-k) + \sum_{l=1}^{L} a_l \frac{\partial y(n-l)}{\partial a_k(n)}$$
$$= y(n-k) + \sum_{l=1}^{L} a_l \beta_k(n-l), \quad k = 0,1,\cdots,L \quad (20)$$

With the derivatives defined in this manner, we have

$$\nabla_{C_N} E_{ms} = -2\varepsilon(n)[\alpha_o(n) \cdots \alpha_M(n)\, \beta_1(n) \cdots \beta_L(n)]^T \quad (21)$$

Now we write the LMS algorithm as follows,

$$C_N(n+1) = C_N(n) - \mu \nabla_C E_{ms} \quad (22)$$

With non-quadratic error surface, we now have a convergence parameter $\mu$ for each $a$ and $b$. We may even wish to have this factor vary with time. Using the current values of the $a$'s in (19) and (20), the LMS algorithm computation for recursive adaptive IIR filter as follows,

$$\begin{cases} y(n) = C_N^T(n) U(n) \\ \alpha_k(n) = x(n-k) + \sum_{l=1}^{L} a_l \alpha_k(n-l), & 0 \le k \le M \\ \beta_k(n) = (n-k) + \sum_{l=1}^{L} a_l \beta_k(n-l), & 1 \le k \le L \\ \nabla_C E_{ms} = -2\varepsilon(n)[\alpha_o(n) \cdots \alpha_M(n)\, \beta_1(n) \cdots \beta_L(n)]^T \\ C_N(n+1) = C_N(n) - \mu \nabla_{C_N} E_{ms} \end{cases} \quad (23)$$

Initialization is the same as in the case adaptive FIR filter except that here, in addition, the $\alpha$'s and $\beta$'s should be set initially to zero unless their values are known.

## 6. Evolutionary Computation: The Genetic Algorithm (GA)

GAs are an evolutionary optimization approach, they are especially appropriate for applications which are vast, nonlinear and potentially discrete in nature. In GA, a population of strings called chromosomes which represent the candidate solutions to an optimization problem is evolved to a better population. It is more common to state the objective of GA as the maximization of some utility or fitness function [21], [22] given by,

$$F(t) = \frac{1}{1+f(t)} \quad (24)$$

where $f(t)$ is the cost function to be minimized. In adaptive filtering, GA operates on a set of filter parameters (the population of chromosomes), in which a fitness values are specified to each individual chromosome. The cost function $f(t)$ in adaptive filtering is taken as the Mean Square Error (MSE) which is given by [4]

$$f(t) = {\varepsilon_j}^2 = \frac{1}{t_e}\sum_{i=1}^{t_e}[d(n) - y_j(n)]^2 \qquad (25)$$

where $t_e$ is the window size over which the errors will be accumulated; $y_j(n)$ is the estimated output associated with the *j-th* set of estimated parameters.

GA consists of main three steps, these are a *selection process*, *crossover*, and *mutation*. The *Selection process* refers to the mechanism of choosing a set of chromosomes from the population that will contribute to the creation of the offsprings for the next generation. The best chromosomes having higher fitness values in the population will get a higher chance of being elected for the next generation, this is how the parents are chosen from the current population. Many methods have been proposed for mate selection in the literature, some of these methods are described in our previous works [21], [22]. On the other hand, *crossover* operation, or mating, is the creation of one or more offspring from the parents selected in the pairing process. The final step of the GA is the *mutation* operator, it is another way of the GA to investigate the cost-surface and it can introduce individuals that never exists in the principal population and preserve the GA from converging too fast before searching the complete cost-surface [23], [24]. The new offsprings (the set of filter coefficients) then form the basis of the next generation. The basic cycle of GA is depicted in Fig. 5.

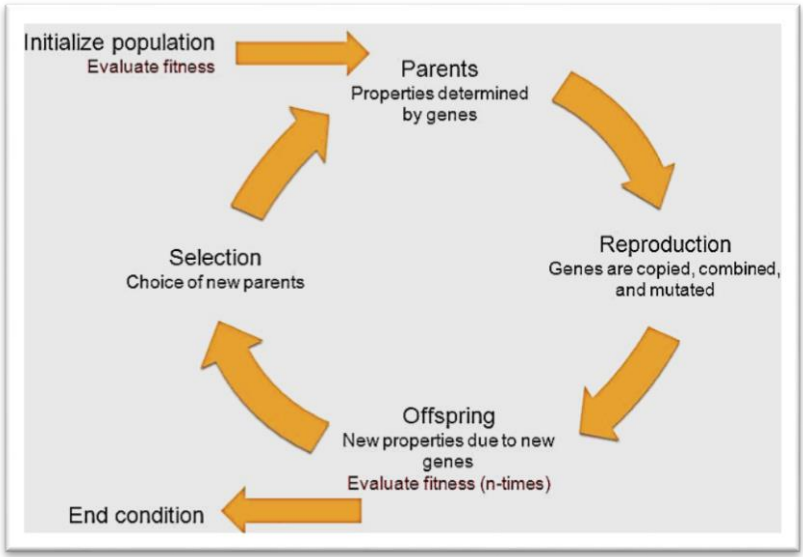

Fig. 5. GA three steps. Selection, crossover, and mutation are involved to create to the evolution from one generation to the next.

Also, the GA will search within each generation for the minimum of the estimation error $\varepsilon_{min} = \min({\varepsilon_j}^2)$ for every chromosome in the entire population with an attempt of driving $\varepsilon_{min}$ to an acceptable minimum value or to zero in the subsequent generations. Adopting the standard GA itself in adaptive filtering leads to a slow convergence rate, this is due to the tremendously big searching space of the GA, which makes the mutation (randomization) process wasting time in examining solutions over improper directions [4].

## 7. The Main Results

7.1. The Effect of Colored Signal On the Adaptation Process of LMS Algorithm

Suppose that the input signal $x(n)$ is passed through a digital Low-Pass Filter (LPF), then, the output of the digital filter is applied to the adaptive system as shown in Fig. 6. To show the effect of the digital low pass filter on the adaptation process, let us discuss the difference between the input and the output signals of the digital filter. The input to the digital filter is shown in Fig. 7(a), where Fig. 7(b) shows the autocorrelation $\phi_{ij}(t,n)$, where $\phi_{ij}(t,n) = x_i(n)\,x_j(n)$, with $i,j = 1,2,\cdots,N-1$, $t = |i-j|$. We notice from Fig. 7 that the input signal $x(n)$ is an impulse signal at each instant and is correlated with itself only and it is never correlated with other impulses. The output of the digital low pass filter is also a random signal as shown in Fig. 8.

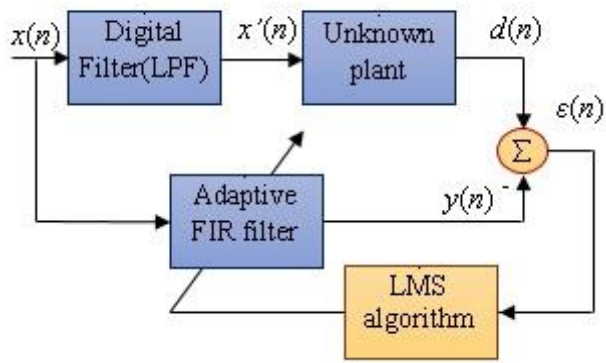

Fig. 6. System identification configuration for the case of the colored input signal.

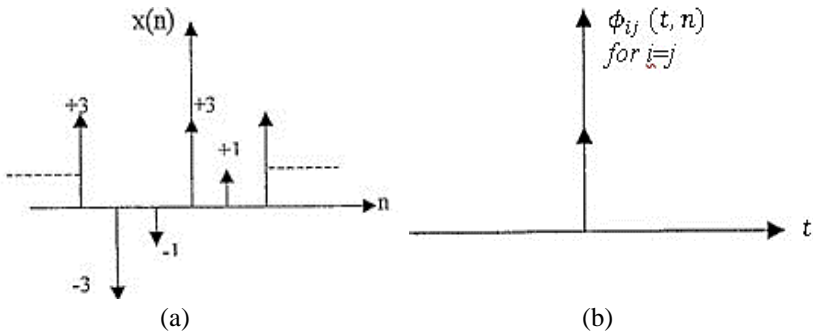

Fig. 7. The input signal and its autocorrelation, (a) the input signal $x(n)$ to the digital LPF, (b) the autocorrelation $\phi_{ij}(t,n)$ of the input signal $x(n)$, where $\phi_{ij}(t,n) = x_i(n)\,x_j(n)$, with $i,j = 1,2,\cdots,N-1, t = |i-j|$.

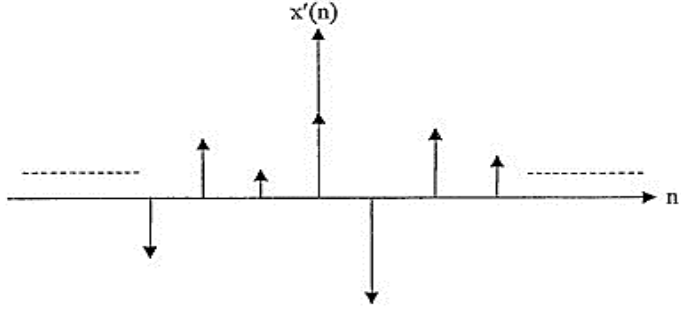

Fig. 8. The output of the digital low pass filter.

The spectral characteristics of the random signal are obtained by computing the Fourier transform for the correlation $\phi_{ij}(t,n)$. of the input signal $x(n)$. The power spectral density of the input signal $x(n)$ is shown in Fig 9., which is flat for all frequencies (White spectrum).

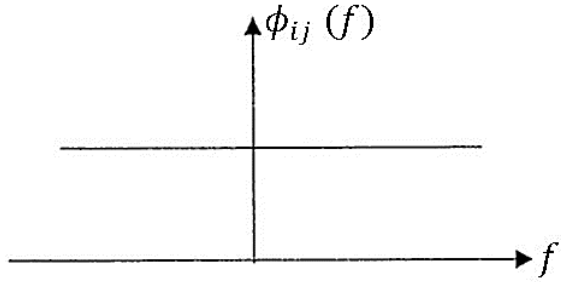

Fig 9. The power spectral density of the input signal $x(n)$.

It is seen from the demonstration of the LMS convergence that the algorithm convergence time $\tau$ is determined as,

$$\tau = \frac{1}{\mu \lambda_{min}} = \frac{1}{2\alpha} \frac{\lambda_{max}}{\lambda_{min}} \tag{26}$$

where $\alpha = \mu \frac{\lambda_{max}}{2}$, $\lambda_{max}$ and $\lambda_{min}$ are the minimum and maximum eigenvalues of the autocorrelation matrix $R_{ms}$ respectively. The ratio of the maximum to minimum eigenvalue is called the *eigenvalue disparity* and determines the speed of convergence. The matrix $R_{ms}$ is given as

$$R_{ms} = \begin{bmatrix} \phi_{00}(0,n) & \phi_{01}(1,n) & \phi_{02}(2,n) & \cdots & \phi_{0\,N-1}(N-1,n) \\ \phi_{10}(1,n) & \phi_{11}(0,n) & \phi_{12}(1,n) & \cdots & \phi_{1\,N-1}(N-2,n) \\ \phi_{20}(2,n) & \phi_{21}(1,n) & \phi_{22}(0,n) & \cdots & \phi_{2\,N-1}(N-3,n) \\ \vdots & \vdots & \vdots & \vdots & \vdots \\ \phi_{N-1\,0}(N-1,n) & \phi_{N-1\,1}(N-2,n) & \phi_{N-1\,2}(N-3,n) & \cdots & \phi_{N-1\,N-1}(0,n) \end{bmatrix} \tag{27}$$

where $\phi_{ij}(t,n) = x_i(n)\, x_j(n)$, with $i,j = 1,2,\cdots,N-1$, $t = |i-j|$. It is seen from (26) that the convergence time is inversely proportional to $\mu$ and depends only on the nature of the input sequence $x(n)$. The physical interpretation of the eigenvalues of $R_{ms}$ can be illustrated by comparing them with the spectrum of the input signal $x(n)$. It is a classical result of Toeplitz form theory that the eigenvalues are bounded by,

$$X(e^{j\omega})_{min} < \lambda_i < X(e^{j\omega})_{max} \quad , \, i = 0,1,\ldots,N-1 \tag{28}$$

where $X(e^{j\omega})$ is the power spectral density of the input or the Fourier transform of the autocorrelation function $\phi_{ij}(t,n)$ (elements of the $R_{ms}$ matrix). As the order of the matrix, $N$, tends to infinity,

$$\begin{cases} \lambda_{min} \to X(e^{j\omega})_{min} \\ \lambda_{max} \to X(e^{j\omega})_{max} \end{cases} \tag{29}$$

Given that the convergence time $\tau$ can be expressed as in (26), we infer that the spectra that with the ratio of the maximum to minimum spectrum is large results in sluggish convergence. Spectra with an eigenvalue disparity near unity (i.e., flat spectra) lead to rapid convergence. The conjecture about such results is that large correlation among the input samples is related to a large eigenvalue disparity, which in turn decelerates the convergence of the adaptation process of the FIR filter.

Now, we show the effect of colored signal $x'(n)$ on the convergence speed of adaptation process. The autocorrelation function $\phi'_{ij}(t,n)$ of the colored signal $x'(n)$ is shown in Fig. 10.

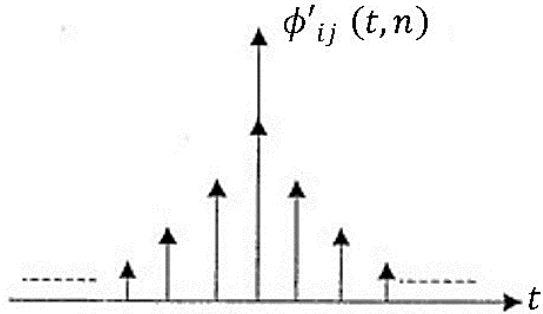

Fig. 10. The autocorrelation of the colored signal $x'(n)$.

From Fig. 10, we can verify that the impulse $x'(0)$ is more correlated with itself and less with other impulses. The spectral density of the input signal can be obtained by computing the Fourier transform of the autocorrelation function (the elements of $R_{ms}$ matrix) as shown in Fig. 11. So that the eigenvalue disparity $\lambda_{max}/\lambda_{min} \to X(e^{j\omega})_{max}/X(e^{j\omega})_{min}$ will be larger than the white signal. Therefore, the effect of the colored signal will result in slow convergence.

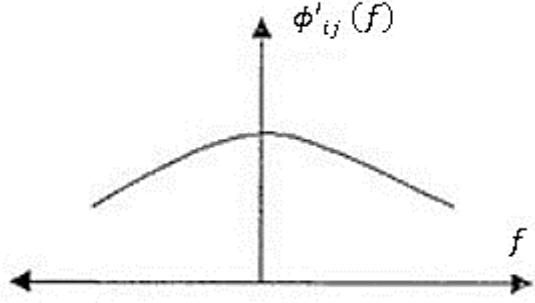

Fig. 11. The power spectral density of the digital filter output $x'(n)$.

7.2. The Integrated LMS Algorithm with Genetic Search Approach (LMS-GA)

As said earlier, GA is slow for tuning in adaptive filtering; while, the gradient-descent techniques behave well in local regions only. To overcome the difficulties of the GA and gradient-descent techniques, we propose in this section a novel technique for the learning the adaptive FIR and IIR filters that incorporates the quintessence and features of both algorithms.

The essential principle of our novel learning algorithm is to combine the evolutionary idea of GA into the gradient-descent technique in order to give an organized random searching amid the gradient-descent calculations. The filter coefficients are represented as a chromosome with a list of real numbers in our proposed technique. The LMS algorithm is embedded in the mutation process of GA to discover the fastest shortcut path in adjusting the optimal solution through the learning process. Each time the LMS learning tool get caught in a local minimum, or the convergence of the LMS algorithm is slow (i.e., the gradient of the error is within a specific range), we begin the GA by arbitrarily varying the estimated filter parameters values to obtain a new sets of filter coefficients. The proposed learning algorithm, namely, LMS-GA, chooses the filter among the new filters and the first one with the smallest MSE (best fitness value) as the new candidate to the next evolution. The above process will be done more and more if the convergence speed is detected to be sluggish at uniform intervals or the LMS algorithm stuck in one more local minimum. In the suggested learning technique, the parameters of the filter are varied during each evolution according to,

$$\Theta_i = \Theta + \sigma_i \mathcal{D} \quad 1 \leq i \leq m \quad (30)$$

where $\mathcal{D}$ is the permissible offset range for each evolution, $m$ is the offsprings number produced in the evolutions, $\Theta_i$ denotes the *i-th* offsprings that are produced by the parents filter $\Theta$, and $\sigma_i \in [-1, +1]$ is a random number. To pick the optimum filter to be the next candidate amongst the sets of new offsprings in the course of each evolution, we calculate the MSE for each new filter ($\Theta_i$, $1 \leq i \leq m$) by (25) for block of time $t_e$. The filter with the smallest MSE will be selected as the next candidate for the subsequent phase of learning process. We represent the behavior of the new proposed LMS-GA by the flowchart shown in Fig. 12. The *ΔE(n)* in the flowchart of Fig. 12 is the error gradient and defined as

$$\Delta E(n) = \frac{e(n) - e(n-\gamma)}{\gamma}$$

where $\gamma$ is the window size for estimation of $\Delta E$

The computational complexity of the LMS algorithm of the FIR filter for the case of the white input signal is found to be $(2N)$ multiplication per iteration, where $N$ is the length of the FIR filter. While the computational complexity required for the IIR filter is equal to $(M + L)(L + 2)$ where $L$ is the backward length and $M$ is the forward length of the IIR filter for the same order of both FIR and IIR filters (i.e., $N = L + M$). The computational complexity of the LMS algorithm of the FIR filter with colored the nput signal is given as $N \cdot P + 2N$, where $P$ is the length of the digital LPF.

## 8. Numerical Results

Consider 4-Tap adaptive FIR filter in channel equalization task and as a plant model for the purpose of system identification,

$$H(z) = 0.03 + 0.24z^{-1} + 0.54z^{-2} + 0.8z^{-3} \tag{31}$$

The basic idea of the system identification using adaptive FIR filtering depends on matching the coefficients of the adaptive filter to that of the plant. The convergence factor $\mu$ regulates the adaptation stability and convergence speed. The results of applying the standard LMS algorithm on adaptive FIR filter to identify the parameters of (31) with different values of $\mu$ are shown in Table 1.

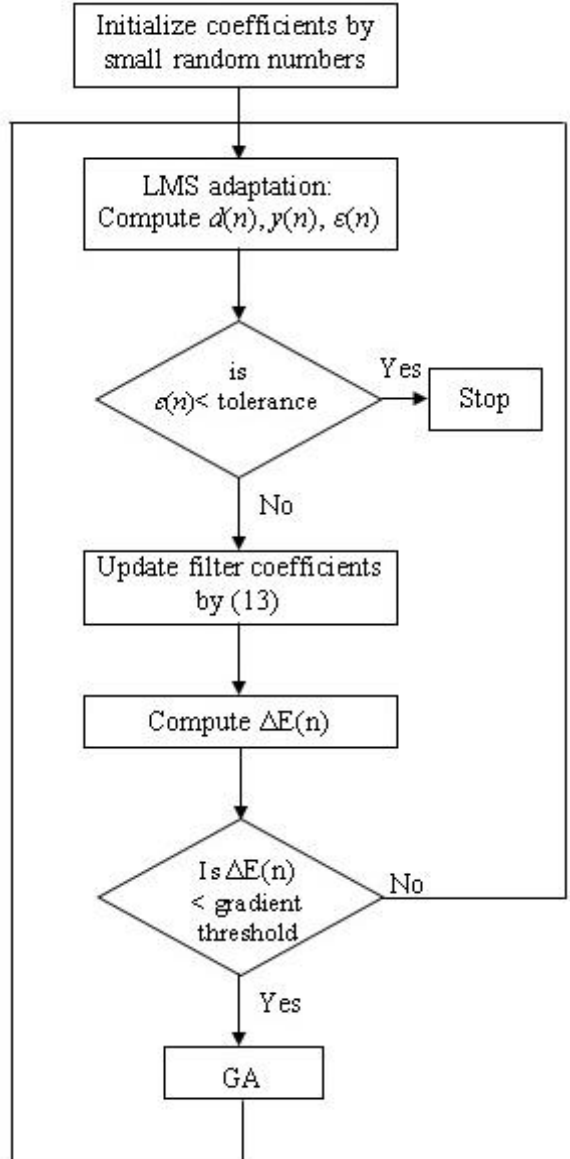

Fig 12. Flowchart of the new learning algorithm (LMS - GA).

Table 1: Simulation results of FIR adaptation (White signal)

| Step size $\mu$ | No. of iterations | MSE /dB | Adaptive filter coefficients | | | |
|---|---|---|---|---|---|---|
| | | | $C_1$ | $C_2$ | $C_3$ | $C_4$ |
| 0.04 | 399 | -167.200 | 0.03 | 0.24 | 0.54 | 0.8 |
| 0.045 | 130 | -166.250 | 0.03 | 0.24 | 0.54 | 0.8 |
| 0.095 | 1911 | -166.278 | 0.03 | 0.24 | 0.54 | 0.8 |

A special class of input signal is generated to train the weights of the adaptive FIR filter, it consists of four level values (-3, -1, +1, and +3) and governed by a uniformly generated random input $R \in [0, 1]$ as shown in Fig. 13, e.g., if the random number is $R_{oo}$, i.e., it is in the range [0, 0.25], then $x(n) = -3$, the same for other values of $R$.

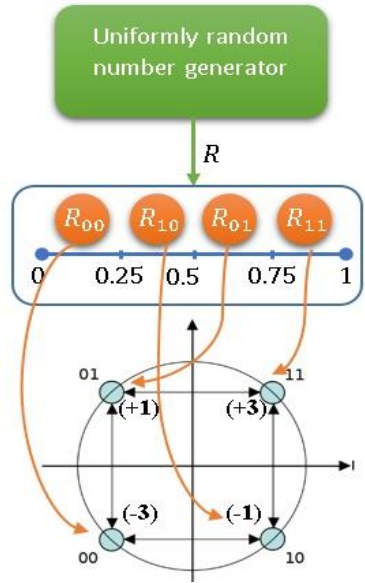

Fig. 13. Four levels input scheme used to train FIR and IIR adaptive filters.

These four level input values generated using the scheme proposed in Fig. 13 are entered repeatedly into the input channel $x(n)$ of Fig. 14 until a convergence is reached or maximum number of iterations are achieved. The learning curves for three different values of $\mu$ are shown in Fig. 15.

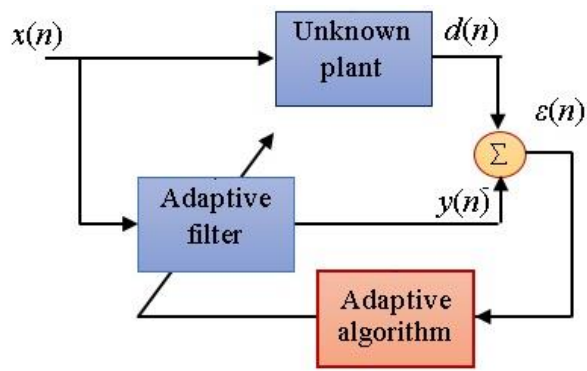

Fig. 14 System identification block diagram.

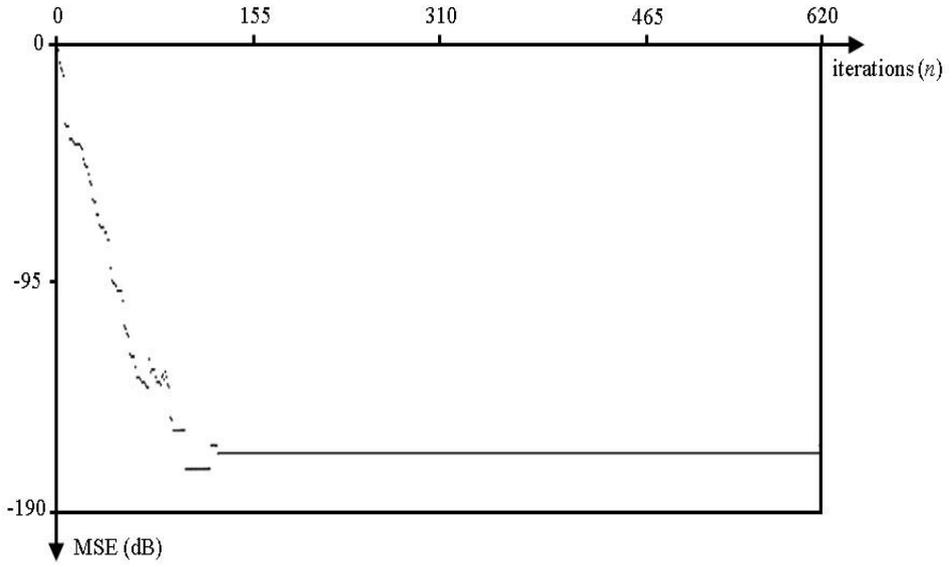

(a)

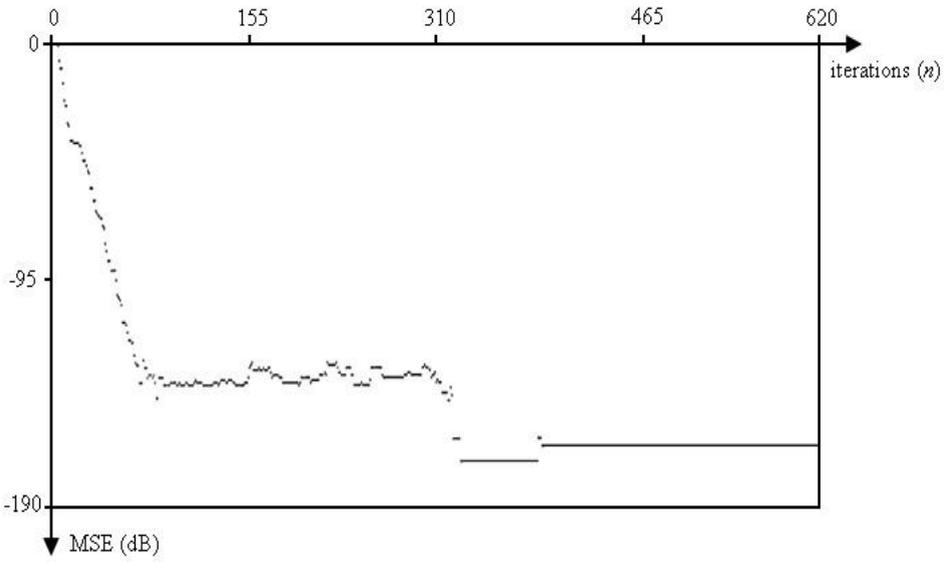

(b)

Fig. 15. Continues…

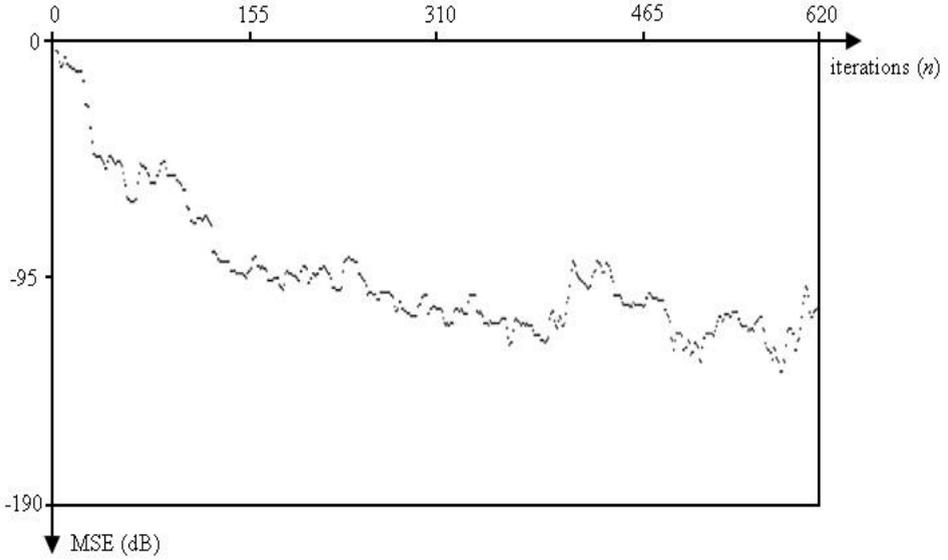

(c)

Fig. 15. The learning curves for FIR adaptive filter with LMS algorithm, (a) $\mu = 0.045$, (b) $\mu = 0.04$, (c) $\mu = 0.095$.

From the above results, we observe that the optimum value of $\mu$ that corresponds to less number of iterations is found to be (0.045). For the case of the coloured input signal, the same input produced using the scheme of Fig. 13 is applied on the input channel of Fig. 6, we conclude that the spectra with the ratio of the maximum to minimum spectrum is large results in sluggish convergence. Spectra with an eigenvalue disparity near unity (i.e., flat spectra) lead to rapid convergence. The digital filter of Fig. 6 used in this simulation is of 8-Tap FIR LPF type given as,

$h(n) = [0.0012654 - 0.0052341 - 0.0019735 - 0.0023009\ 0.022366\ 0.12833\ 0.0013\ 0.0012]$;

While the plant dynamics is given in (31). Different $\mu$'s have been used in coloured input signal case study with the results given in Table 2. We note that the optimum value of $\mu$ that corresponds to the less number of iterations is found to be 3. The learning curves for different values of $\mu$ are illustrated in Fig. 16. The best value of $\mu$ is found with MSE of -84.21901 dB.

Table 2: Simulation results of FIR adaptation (Coloured signal)

| Step size $\mu$ | No. of iterations | MSE /dB | Adaptive filter coefficients | | | |
|---|---|---|---|---|---|---|
| | | | $C_1$ | $C_2$ | $C_3$ | $C_4$ |
| 0.9 | 2320 | -135.591 | 0.03 | 0.24 | 0.54 | 0.8 |
| 3 | 358 | -163.131 | 0.03 | 0.24 | 0.54 | 0.8 |
| 4 | 362 | -173.604 | 0.03 | 0.24 | 0.54 | 0.8 |

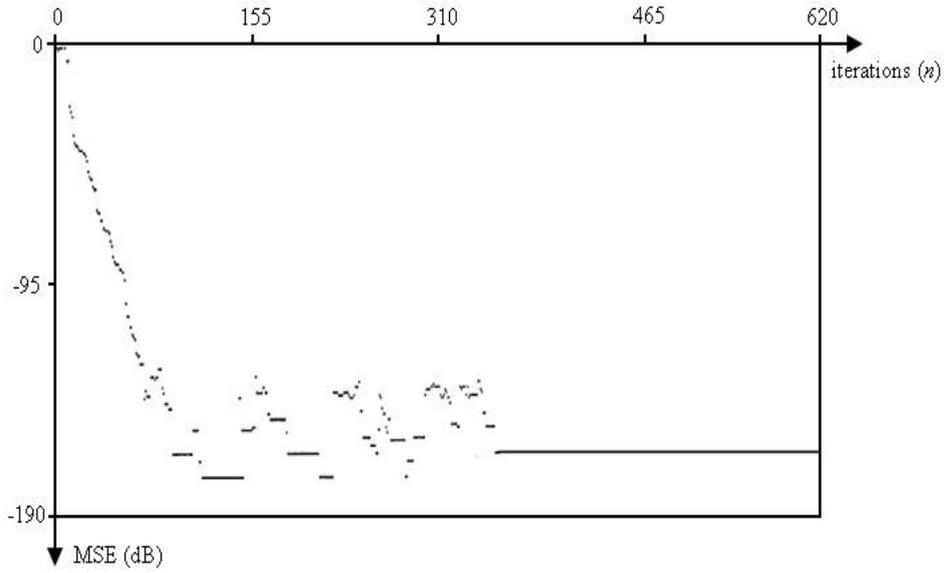

(a)

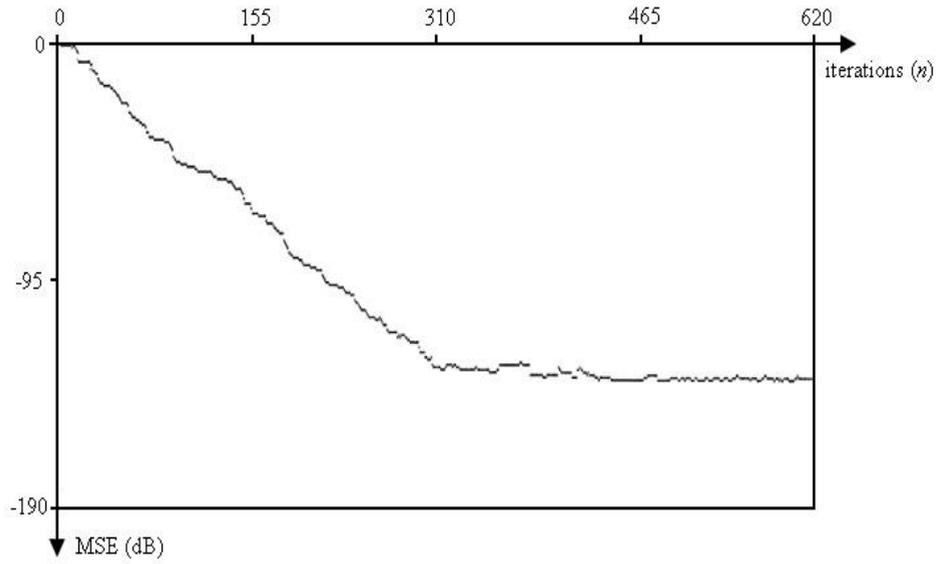

(b)

Fig. 16. Continues…

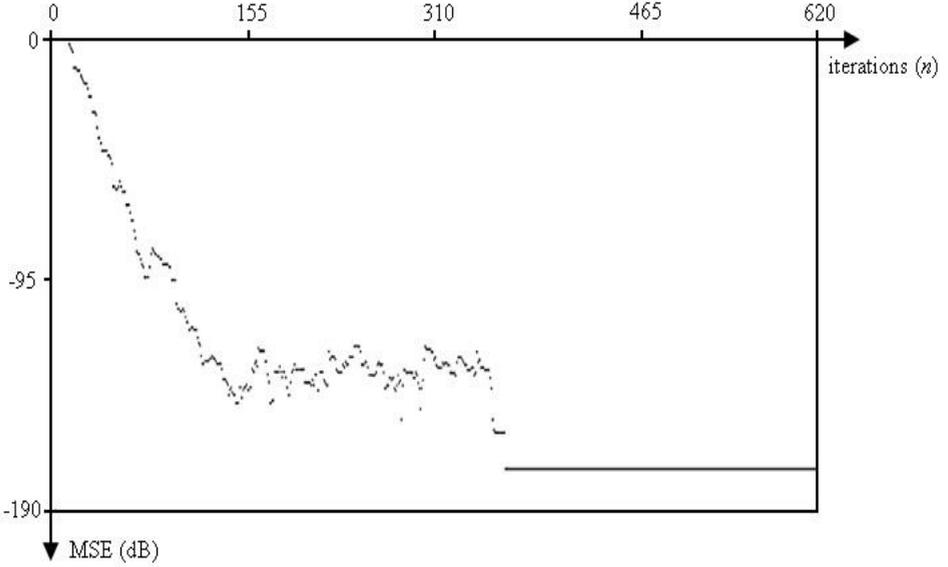

(c)

Fig. 16. The learnings curves for the case of colored input signal, (a) $\mu = 3$, (b) $\mu = 0.9$, (c) $\mu = 4$.

Concerning the adaptive IIR filter, the error surface is generally a multi-modal against filter parameters. The adaptation techniques for the case of adaptive IIR filter can easily get stuck at a local minimum and escape away from the global minimum. Some of the adaptive IIR filter coefficients will be matched with that of the plant and the other will be constant at certain values, which means that these coefficients are stuck at local minima. The following 1st order transfer function $H(z)$ is used to represent the plant,

$$H(z) = \frac{0.6}{1 - 0.2z^{-1}} \tag{32}$$

The results of the system identification using IIR adaptive filtering are shown in Table 3. As can be seen, that the best value of $\mu$ was found to be 0.065. The IIR error surface here is a special case as it is an uni-modal (local minimum does not exists) and having a global optimum only. However, the practical problem still exists here which is the pole of the adaptive filter may move outside of the unit circle resulting in an unstable system. To solve this problem we use a certain criterion that states when the magnitude of the pole exceeds unity, we limit its magnitude to be less than one. The learning curves for different values of $\mu$ are shown in Fig. 17.

Table 3: Simulation of adaptive IIR filter (white signal)

| Step size $\mu$ | No. of iterations | MSE /dB | Adaptive filter coefficients | |
|---|---|---|---|---|
| | | | a | b |
| 0.04 | 142 | -157.373 | -0.2 | 0.6 |
| 0.06 | 63 | -143.939 | -0.2 | 0.6 |
| 0.1 | 134 | -174.030 | -0.2 | 0.6 |

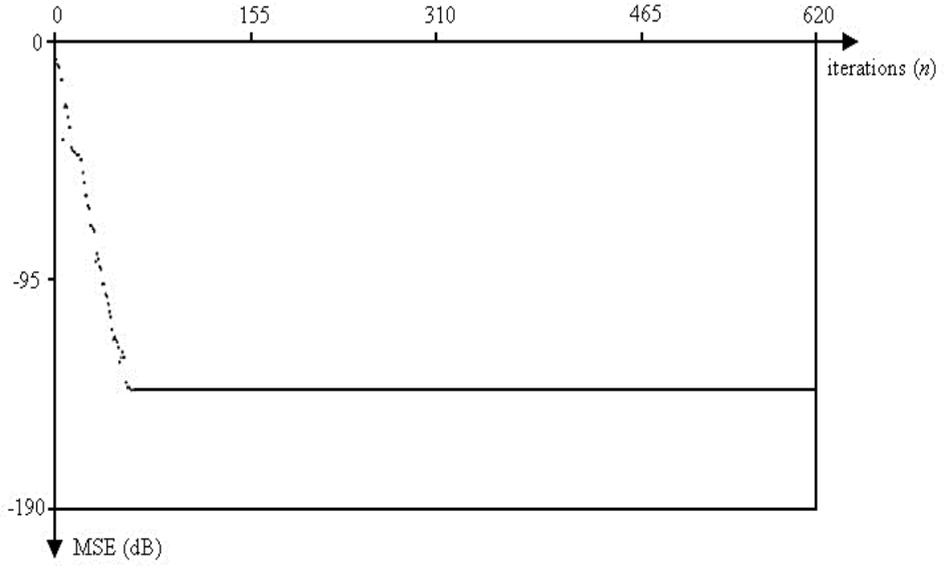

(a)

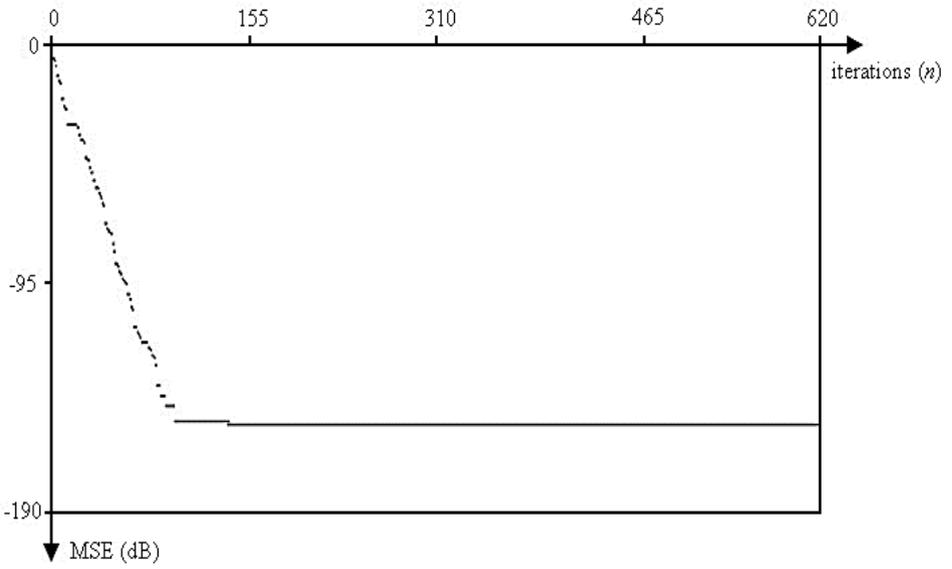

(b)

Fig. 17 Continues…

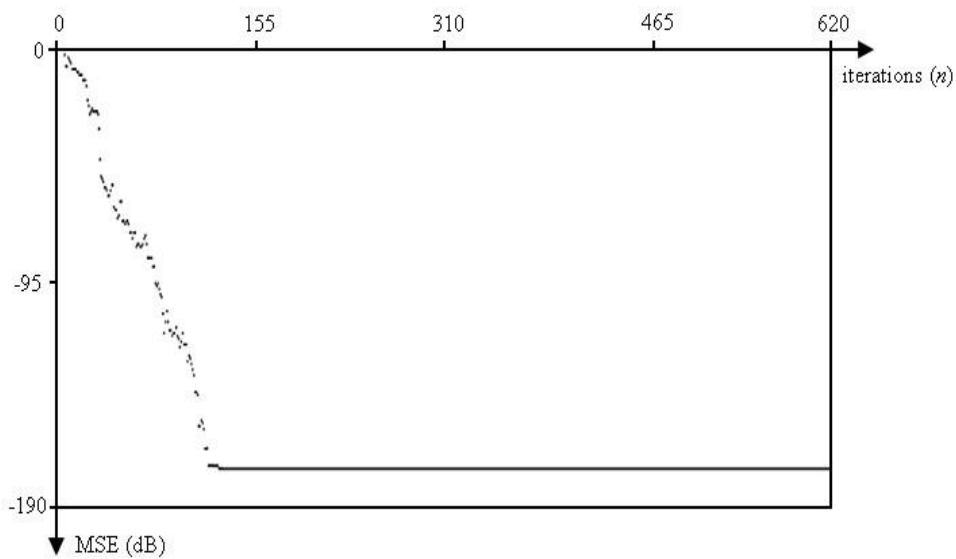

(c)

Fig. 17. The learnings curves for the case of IIR adaptive filter, (a) $\mu = 0.06$, (b) $\mu = 0.04$, (c) $\mu = 0.1$.

The new LMS-GA learning tool can be applied to the system identification as shown in Fig. 14 with FIR adaptive filter instead of IIR. Then, we can deduce the learning curve with windows size 8 and offsprings $m = 5$ and offset $D = 0.02$ as shown in Fig. 18. One can see that the pure LMS algorithm is faster than the new learning algorithm because the LMS-GA is a random technique which is applied to multi-modal error surface. In the case of unimodal error surface ( as the case of FIR adaptive filter), the pure LMS algorithm is the better choice than other algorithms.

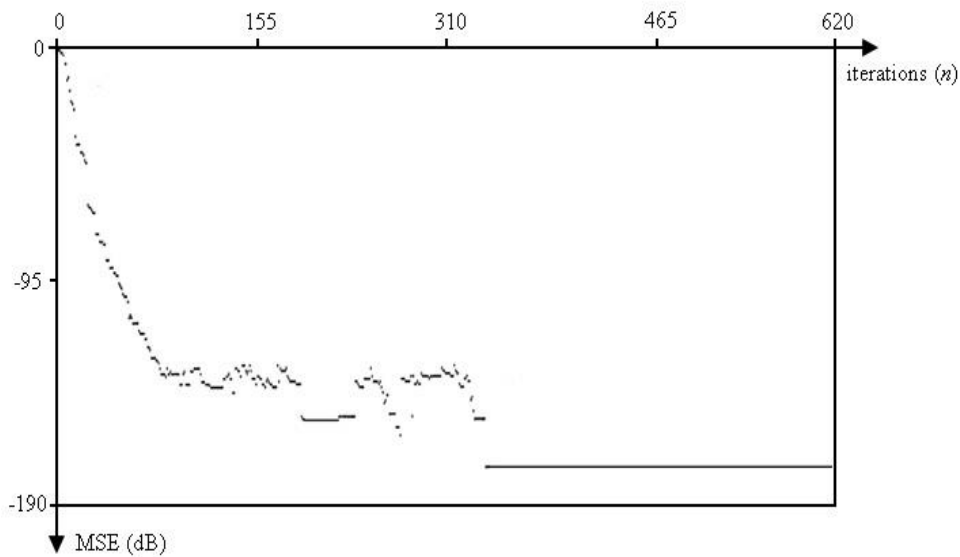

Fig. 18. The learning curve of FIR adaptive filter with integrated LMS-GA.

For the sake of comparison study, the LMS-GA learning tool can be applied to the same configuration of adaptive system identification shown in Fig. 14 with the results listed in Table 4.

Table 4: Simulation results of both standard LMS and LMS-GA algorithms

|  |  | Standard LMS algorithm with optimum $\mu$ | LMS-GA with optimum $\mu$ |
|---|---|---|---|
| No. of iterations |  | 130 | 336 |
| MSE (dB) |  | -166.250 | -173.604 |
| Filter Coefficients | $C_1$ | 0.03 | 0.03 |
|  | $C_2$ | 0.24 | 0.24 |
|  | $C_3$ | 0.54 | 0.54 |
|  | $C_4$ | 0.8 | 0.8 |

The proposed LMS-GA learning tool is exploited to learn an adaptive IIR filter to recover the performance of the gradient descent technique (e.g., the LMS algorithm) with multi-modal error surface. To compare the new LMS-GA learning tool with the standard LMS algorithm, we must determine the window size δ the window for estimation of ΔE and the Gradient Threshold (GT), these can be calculated from the learning curve of the pure LMS algorithm as follows, window size τ is calculated as being the number of iterations between the first iteration and the iteration at which the learning curve fluctuate with a small variations. GT is calculated by determining the maximum and minimum values of these fluctuations. Now, GT can be determined as, GT=(max.swing-min.swing)/ δ. If these two parameters are calculated, we can apply the procedure of the new learning algorithm of Fig. 12 on adaptive IIR filter of a unimodal error surface as in (32). We conclude that the new learning algorithm will converge to the same MSE (the same MSE the pure LMS reached to it) but with a fewer number of iterations as shown in Fig. 19.

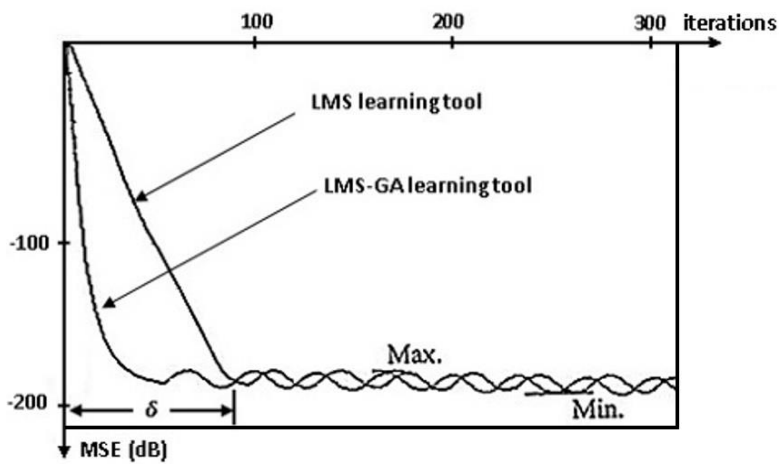

Fig. 19 The convergence performance of the LMS-GA tool for adaptive IIR filter.

## 9. Conclusions

In this work, adaptive algorithms are adopted to learn the parameters of the digital FIR and IIR filters such that the error signal is minimized. These algorithms are the standard LMS algorithm and the LMS-GA one. The numerical instabilities inherent in other adaptive techniques do not exist in the standard LMS algorithm. Moreover, prior information of the signal statistics is not required, i.e., the autocorrelation and cross-correlation matrices. The LMS algorithm produces only estimated adaptive filter coefficients. These estimated coefficients match that of the plant progressively through the time as the coefficients are changed and the adaptive filter learns the signal characteristics, then identify the underlying plant. Due to the multimodality of the error surface of adaptive IIR filters, a new learning algorithm, namely, LMS-GA is proposed in this paper which integrates the genetic searching methodology LMS algorithm and speeds up the adaptation procedure and offers universal searching ability. Besides, the LMS-GA preserved the characteristics and the simplicity of the standard LMS learning algorithm and it entails comparatively fewer computations and had a fast convergence rate as compared to the standard GA. The numerical simulations evidently elucidated that the LMS-GA outperforms the standard LMS in terms of the capability to determine the global optimum solution and the faster convergence rate to this solution.

## References

[1]     S. D. S. Bernard Widrow, *Adaptive Signal Processing*. Prentice-Hall, 1985.
[2]     J. C. S. and J. V. J. Gerardo Avalos, "Applications of Adaptive Filtering," in *Adaptive Filtering Applications*, Lino Garcia Morales, Ed. InTech, 2011, pp. 1–20.
[3]     M. Shams Esfand Abadi, H. Mesgarani, and S. M. Khademiyan, "The wavelet transform-domain LMS adaptive filter employing dynamic selection of subband-coefficients," *Digit. Signal Process.*, vol. 69, pp. 94–105, 2017.
[4]     C. Y. C. S.C. Ng, S.H. Leung, "The Genetic Search Approach: A new Learning Algorithm for Adaptive IIR Filtering," *IEEE Signal Process. Mag.*, vol. 13, no. 6, pp. 38–46, 1996.
[5]     S. M. Kumpati S. Narendra, "Neural Networks for System Identification," in *IFAC System Identification*, vol. 30, no. 11, pp. 735–742.
[6]     D. R. Santosh Kumar Behera, "System Identification Using Recurrent Neural Network," *Int. J. Adv. Res. Electr. Electron. Instrum. Eng.*, vol. 3, no. 3, pp. 8111–8117, 2014.
[7]     H. Jaeger, "Adaptive nonlinear system identification with echo state networks," in *Advances in Neural Information Processing Systems*, 2003, pp. 593–600.
[8]     W. Zhang, "System Identification Based on a Generalized ADALINE Neural Network," in *American Control Conference (ACC)*, 2007, vol. 11, no. 1, pp. 4792–4797.
[9]     I. K. Ibraheem, "System Identification of Thermal Process using Elman Neural Networks with No Prior Knowledge of System Dynamics," *Int. J. Comput. Appl. (0975*, vol. 161, no. 11, pp. 38–46, 2017.
[10]    V. Katari, S. Malireddi, S. K. S. Bendapudi, and G. Panda, "Adaptive nonlinear system identification using comprehensive learning PSO," in *Communications, Control and Signal Processing, 2008. ISCCSP 2008. 3rd International Symposium on*, 2008, no. March, pp. 434–439.
[11]    A. C. Sinha, Rashmi, "Adaptive Filtering Via Wind Driven Optimization Technique," in *3rd IEEE International Conference on "Computational Intelligence and Communication Technology" (CICT)*, 2017, pp. 1–5.
[12]    Q. L. Qian Zhang, Sa Wu, "A PSO identification algorithm for temperature adaptive adjustment system," in *IEEE International Conference on Industrial Engineering and Engineering Management (IEEM)*, vol. 2016–Janua, pp. 752–755.
[13]    J. Zhang and P. Xia, "An improved PSO algorithm for parameter identification of nonlinear dynamic hysteretic models," *J. Sound Vib.*, vol. 389, pp. 153–167, 2017.
[14]    A. Sarangi, S. K. Sarangi, M. Mukherjee, and S. P. Panigrahi, "System identification by Crazy-cat swarm optimization," in *2015 International Conference on Microwave, Optical and Communication Engineering (ICMOCE)*, 2015, pp. 439–442.
[15]     and K. K. A.-M. Thamer M. Jamel, "SIMPLE VARIABLE STEP SIZE LMS ALGORITHM FOR ADAPTIVE IDENTIFICATION OF IIR FILTERING SYSTEM," in *The 5th International Conference*

| | |
|---|---|
| | *on Communications, Computers and Applications (MIC-CCA)*, pp. 23–28. |
| [16] | S. A. Ghauri and M. F. Sohail, "System identification using LMS, NLMS and RLS," in *Proceeding - 2013 IEEE Student Conference on Research and Development, (SCOReD)*, no. December, pp. 65–69. |
| [17] | L. Lu and H. Zhao, "A novel convex combination of LMS adaptive filter for system identification," in *12th International Conference on Signal Processing Proceedings (ICSP)*, vol. 2015–Janua, no. October, pp. 225–229. |
| [18] | R. Yu, Y. Song, and M. Nambiar, "Fast system identification using prominent subspace LMS," *Digit. Signal Process. A Rev. J.*, vol. 27, no. 1, pp. 44–56, 2014. |
| [19] | F. Titel and K. Belarbi, "Identification of Dynamic systems using a Genetic Algorithm-based Fuzzy Wavelet Neural Network approach," in *Proceedings of the 3rd International Conference on Systems and Control*, 2013, pp. 6–11. |
| [20] | N. A. S. Alwan, "Time Varying Channel Estimation Based on Adaptive Filtering Techniques," Baghdad University, 1991. |
| [21] | A. A. A. Ibraheem Kasim Ibraheem, "Application of an Evolutionary Optimization Technique to Routing in Mobile Wireless Networks Application of an Evolutionary Optimization Technique to Routing in Mobile Wireless Networks," *Int. J. Comput. Appl.*, vol. 99, no. 7, pp. 24–31, 2014. |
| [22] | I. K. Ibraheem and A. A. A. Ibraheem Kasim Ibraheem, "Design of a Double-objective QoS Routing in Dynamic Wireless Networks using Evolutionary Adaptive Genetic Algorithm," *Int. J. Adv. Res. Comput. Commun. Eng.*, vol. 4, no. 9, pp. 156–165, 2015. |
| [23] | T. Lu and J. Zhu, "Genetic algorithm for energy-efficient QoS multicast routing," *IEEE Commun. Lett.*, vol. 17, no. 1, pp. 31–34, 2013. |
| [24] | C. W. Ahn and R. S. Ramakrishna, "A genetic algorithm for shortest path routing problem and the sizing of populations," *IEEE Trans. Evol. Comput.*, vol. 6, no. 6, pp. 566–579, 2002. |